\documentclass[journal]{IEEEtran}
\usepackage[boxed]{algorithm}
\usepackage{algorithmic}
\usepackage{enumitem}
\usepackage{graphics}
\usepackage[dvips]{epsfig}
\usepackage{amsmath,amssymb,times,url}
\usepackage[tight]{subfigure}
\usepackage{balance}

\begin{document}

\title{Performance Rescaling of Complex Networks}
\author{Constantinos~Psomas,~Fragkiskos~Papadopoulos
\thanks{\copyright 2013 IEEE. Personal use of this material is permitted. Permission from IEEE must be obtained for all other uses, including reprinting/republishing this material for advertising or promotional purposes, collecting new collected works for resale or redistribution to servers or lists, or reuse of any copyrighted component of this work in other works.}
\thanks{The authors are with the Department of Electrical Engineering, Computer Engineering and Informatics, Cyprus University of Technology, 33 Saripolou Street, 3036 Limassol, Cyprus (e-mail: c.psomas@cut.ac.cy; f.papadopoulos@cut.ac.cy).}}

\maketitle

\begin{abstract}
Recent progress in network topology modeling~\cite{MaHu07,tomacs_rescale} has shown that it is possible to create smaller-scale replicas
of large complex networks, like the Internet, while simultaneously preserving several important topological properties. However, the constructed replicas do not include notions of capacities and latencies, and the fundamental question of whether smaller networks can reproduce the performance of larger networks remains unanswered. We address this question in this letter, and show that it is possible to predict the performance of larger networks from smaller replicas, as long as the right link capacities and propagation delays are assigned to the replica's links. Our procedure is inspired by techniques introduced in~\cite{tomacs_rescale} and combines a time-downscaling argument from~\cite{shrink}. We show that significant computational savings can be achieved when simulating smaller-scale replicas with TCP and UDP traffic, with simulation times being reduced by up to two orders of magnitude.
\end{abstract}

\section{Introduction}
\label{sec:intro}

Understanding the performance of large-scale complex networks like the Internet~\cite{Dorogovtsev10-book}, and predicting their behavior under new algorithms, protocols, architectures and load conditions, are important research problems.\footnote{By {\em complex (or scale-free) networks\/} we mean here real networks with distributions $P(k)$ of node degrees $k$ following power laws $P(k)\sim k^{-\gamma}$~\cite{Dorogovtsev10-book}.}
A commonly accepted practice is to use simulations for testing and evaluating the performance of such networks. Unfortunately, however, it is often very expensive and inefficient to accurately run large-scale simulations (e.g., with several thousands of nodes), which incorporate realistic traffic and topology models, since the memory and CPU requirements of such simulations seem to be well beyond the reach of available hardware (cf. Section~\ref{sec:performance}). This problem has motivated earlier research involving network topology modeling, particularly Internet topology, in an attempt to find techniques for constructing realistic smaller-scale replicas of given real networks. The most relevant earlier results to our work are the groundbreaking results in~\cite{MaKrFaVa06}, and its extensions~\cite{MaHu07,tomacs_rescale}, which are reviewed below. 

It has been shown in~\cite{MaKrFaVa06} that the Autonomous Systems (AS) Internet topology can be well characterized by its joint degree distribution $p(k,k')$, i.e., its $2K$-distribution---the probability that a link connects nodes (ASs) of degrees $k$ and $k'$. Therefore, if one constructs a synthetic network with the same $p(k,k')$, then this network, called a $2K$-random graph, will have approximately the same global structure as the AS Internet. That is, a large number of topological properties of the original network are well preserved in the synthetic replica, such as degree and distance distributions, assortativity, and other~\cite{MaKrFaVa06}. Once the $p(k,k')$ distribution is preserved one can also reproduce the amount of clustering of the original network, following a $p(k,k')$-preserving clustering-targeting link rewiring procedure as in~\cite{pol_clustering} (see Section~\ref{sec:rescaling_method}). The approach in~\cite{MaKrFaVa06} has been extended in~\cite{MaHu07,tomacs_rescale} for generating topologies of different sizes, with approximately the same $p(k, k')$. It has been also suggested that $2K$-random graphs could provide appropriate descriptions of other observed networks in a variety of settings~\cite{MaKrFaVa06}. However, even though it was shown that \emph{network structure} can be preserved, the question of whether \emph{network performance} can be preserved remains unanswered. Here we address this question, and show that performance can be preserved in $2K$-random replicas, \emph{as long as the right link capacities and propagation delays are assigned to the replica's links.}  More importantly, we show that performance can be preserved even in downscaled replicas, consisting of a significantly smaller number of nodes compared to the original network.  

Unrelated to the work in~\cite{MaHu07,tomacs_rescale,MaKrFaVa06}, another important result that we use in this letter is the time-downscaling law from~\cite{shrink}. Consider a network with a set of link capacities $\{C_i\}$ and a set of propagation delays $\{P_i\}$, where network flows (e.g., TCP or UDP flows) arrive according to a Poisson process. Let $\{\lambda_i\}$ be the set of arrival rates of these flows. Note that while flow arrival times are Poisson, packet arrivals within each flow can arrive according to \emph{any} process, e.g., dictated by TCP dynamics, etc. Let $\alpha \leq 1$ be a scaling factor and do the following operations to construct a \emph{time-stretched} replica: 1) sample each incoming flow independently with probability $\alpha$; 2) reduce link capacities by the same factor $\alpha$; 3) increase propagation delays by a factor $1/\alpha$; and 4) increase protocol timeouts by the same factor $1/\alpha$. In summary, flow arrival rates $\{\lambda_i\}$, capacities  $\{C_i\}$, and propagation delays $\{P_i\}$ change to $\{\alpha \lambda_i\}$, $\{\alpha C_i\}$, and $\{P_i/\alpha\}$, while flow arrival times remain Poisson. 
In simple words, the only difference between the original and scaled system is that the latter runs slower by the factor $\alpha$. Thus, distributions of performance metrics, e.g., queue length distributions and normalized delay distributions, are preserved~\cite{shrink}. (By normalized delays we mean packet or flow delays multiplied by $\alpha$.) We call this result the \emph{time-downscaling} law.  

In the next section, we present our network downscaling procedure and apply it to the AS Internet. In Section~\ref{sec:performance}, we show that the performance of larger networks can be preserved in downscaled replicas produced by our method, and in Section~\ref{sec:conclusion}, we conclude with future research directions.

\section{Rescaling Link Correlations}
\label{sec:rescaling_method}

Consider a scale-free network consisting of $N$ nodes and $L$ links, where each link $i=1,\ldots, L$ has capacity $C_i$, propagation delay $P_i$, and incident nodes with degrees $k_i$ and $k'_i$. Thus, each link $i$ is characterized by the vector $\mathbf{L}_i=(k_i, k'_i, C_i, P_i)$. Let's consider a general setting, where $C_i, P_i$ are possibly correlated with each other and with the degrees $k_i$ and  $k'_i$.
Our approach to create synthetic replicas resembling the original network is similar to \cite{MaKrFaVa06}, with the difference that instead of preserving only the joint degree distribution $p(k, k')$, we preserve the joint distribution $p(k, k', C, P)$, which is the probability that a link connects nodes of degrees $k$ and $k'$, and has capacity $C$ and propagation delay $P$. Note that from $p(k, k', C, P)$ we can obtain the degree distribution $P(k)=(\bar{k}/k)\sum_{k',C,P}P(k,k', C, P)$, where $\bar{k}$ is the average node degree, and $p(k, k')=\sum_{C,P}P(k,k', C, P)$. In this way, we simultaneously ensure that the global structure of the original network is well preserved, and that each network flow has the same probability (as in the original network) of traversing a path of some length $h$, consisting of a sequence of links with vectors $\mathbf{L}_1, \mathbf{L}_2,\ldots, \mathbf{L}_h$. Since link capacity and propagation delay correlations are preserved in every network path, we expect the performance of the synthetic replica to be the same to that of the original system.

Following this approach, we show how to create performance-preserving replicas consisting of $N'=\alpha N$ nodes, where $\alpha \leq 1$ is a downscaling factor. The reason we can do this, is because several important topological characteristics of scale-free networks, including the degree and distance distributions, do not change significantly with their size~\cite{Dorogovtsev10-book}. For example, shortest path lengths grow extremely slowly as $\sim \ln{\ln{N}}$, while
the degree distribution remains power law  with the same $\gamma$, $P(k)\sim k^{-\gamma}$, $k \in [1 \ldots k_{max}]$, but with different $k_{max} \sim N^{\frac{1}{\gamma-1}}$~\cite{Dorogovtsev10-book}. Our procedure is shown in Figure~\ref{fig:rescale_method}, and has been inspired by techniques introduced in~\cite{tomacs_rescale}. In summary, we first compute the empirical distributions (CCDFs) of node degrees $k$, link capacities $C$ and propagation delays $P$ in the original network, and then fit them with smoothing splines (smooth continuous curves) $S_k$, $S_C$, $S_P$ using the \texttt{smooth.spline} method of the R project~\cite{rproject}. Note that spline smoothing can extrapolate the shape of an empirical function beyond the original data range~\cite{tomacs_rescale}. From the distributions $S_k$, $S_C$, $S_P$, we sample $N'$ degree values, and $L'$ capacity and propagation delay values, where $N'$ and $L'$ are the target number of nodes and links respectively, see steps~1-3 in Fig.~\ref{fig:rescale_method}. We then combine these values together according to the correlation profile of  $k, k', C, P$ in the original network, to build the target synthetic network. Our procedure is centered around matching the sample ranks of joint node degrees, capacities, and propagation delays from a set of sampled edges of the original network to the set of edges in the rescaled network, see steps~6-9 in Fig.~\ref{fig:rescale_method}. For an example of how the method works one can assign values to variables $N'$, $k^a_1, k^a_2, \dots, k^a_{N'}, C^a_1, C^a_2,\dots,C^a_{L'}, P^a_1, P^a_2, \dots, P^a_{L'}$ in steps~2, 3 of  Fig.~\ref{fig:rescale_method}, and to vectors $\mathbf{L}_1, \mathbf{L}_2,\ldots, \mathbf{L}_{L'}$ in step~6. The code implementing the procedure can be found online at~\cite{rescaling_code}.

\begin{figure}[!ht]
\begin{center}
\begin{minipage}{3.7in}
\begin{algorithm}[H]
\begin{algorithmic}[1]
\small
\REQUIRE Original network topology with link capacities and propagation delays;
\REQUIRE Size $N'$ of the target synthetic topology;
~\\
// \sffamily Construct marginals.
\normalfont
\begin{itemize}[leftmargin=0.1in]
\item[1:] Approximate the empirical distributions of node degrees $k$, capacities $C$, and propagation delays $P$ of the original network with smoothing splines
$S_k$, $S_C$, $S_P$, respectively;
\item[2:] Sample $N'$ values $k^a_i$, $i=1,\ldots,N'$, with probability distribution given by $S_k$, and compute $L'=\frac{1}{2}\sum_{i=1}^{N'}k^a_i$;
\item[3:] Sample $L'$ values $C^a_i$ and $L'$ values $P^a_i$, $i=1,\ldots,L'$, with probability distribution given by $S_C$ and $S_P$ respectively;
\item[4:] Let $k^a_i$ be the target degree of node with id $i$, then node $i$ has $k^a_i$ stubs (edge-ends) attached to it. Label each of these stubs with $k^a_i$;
\item[5:] Let $R_k^{a}$ be the list of all stub labels, $R_C^{a}$ be the list of $C^a_i$, and $R_P^{a}$ be the list of $P^a_i$, all sorted in the non-decreasing order of values;
\end{itemize}
// \sffamily Construct correlations (merge marginals).
\normalfont
\begin{itemize}[leftmargin=0.1in]
\item[6:]  Sample $L'$ links from the original network and denote their vectors by $\mathbf{L}_i=(k^b_i, k'^b_i, C^b_i, P^b_i)$, $i=1,\ldots,L'$;
\item[7:] Let $R_{k}^{b}$ be the list of stub labels $k^b_i, k'^b_i$  taken from the vectors $\mathbf{L}_i$, $i=1,\ldots,L'$. Similarly, let $R^b_C$ and $R^b_P$ be the lists of $C^b_i$ and $P^b_i$ respectively. Sort the values of each list in the non-decreasing order;
\item[8:] Let $r(x, R_x)$ be a rank function that returns the position of $x$ in $R_x$. Also, let  $R_x[q]$ denote the value in the $q^{th}$ position of $R_x$;
\item[9:] For each $\mathbf{L}_i=(k^b_i, k'^b_i, C^b_i, P^b_i)$ compute the link vector $\mathbf{L}_i'=$\\$(R_{k}^{a}[r(k^b_i, R_{k}^{b})], R_{k}^{a}[r(k'^b_i, R_{k}^{b})], R_C^{a}[r(C^b_i,R_C^{b})], R_P^{a}[r(P^b_i, R_P^{b})])$, where $R_k^{a}, R_C^{a}, R_P^{a}$ given in step 5;
\end{itemize}
// \sffamily Build the network.
\normalfont
\begin{itemize}[leftmargin=0.1in]
\item[10:]  In each $\mathbf{L}_i'$ replace the node degrees with the corresponding node ids to construct a network link;
\end{itemize}
\ENSURE Network of size $N'$ with link capacities and propagation delays.
\normalfont
\end{algorithmic}
\end{algorithm}
\end{minipage}
\end{center}
\caption{Topology rescaling procedure.}
\label{fig:rescale_method}
\end{figure}

\textbf{Validation.} To verify our procedure we use the AS Internet topology of December 2010, available at~\cite{as_topo_data}. The topology consists of $N=29333$ nodes and has a power law degree distribution with exponent $\gamma=2.1$. We assign capacity and propagation delay values to each link of the topology according to two different scenarios, which we also consider in Section~\ref{sec:performance}. In Scenario 1:  $C(k, k')=\min(k, k')$~Mb/s and $P(k, k')=500\sqrt{kk'}/{k_{max}}$~ms; and in Scenario 2:  $C(k, k')= \min(2000/k, 2000/k')$~Mb/s and $P(k, k')$ uniformly distributed in $[1\ldots 500]$~ms. Scenario 1 represents a case where link capacities and propagation delays are correlated with the degrees $k, k'$ of the nodes the link connects, and where link capacities increase with node degrees. Scenario 2 represents a different case, where link capacities decrease with node degrees, and where propagation delays are not correlated with $k, k'$. We then build downscaled replicas with our procedure, consisting of $N'=7333, 2933$ nodes, i.e., using $\alpha=1/4, 1/10$, and compare their capacity and propagation delay correlation characteristics to those of the original network.  

\begin{figure*}
\centerline{
\subfigure[Capacity correlations.]{\includegraphics [width=1.8in, height=1.31in]{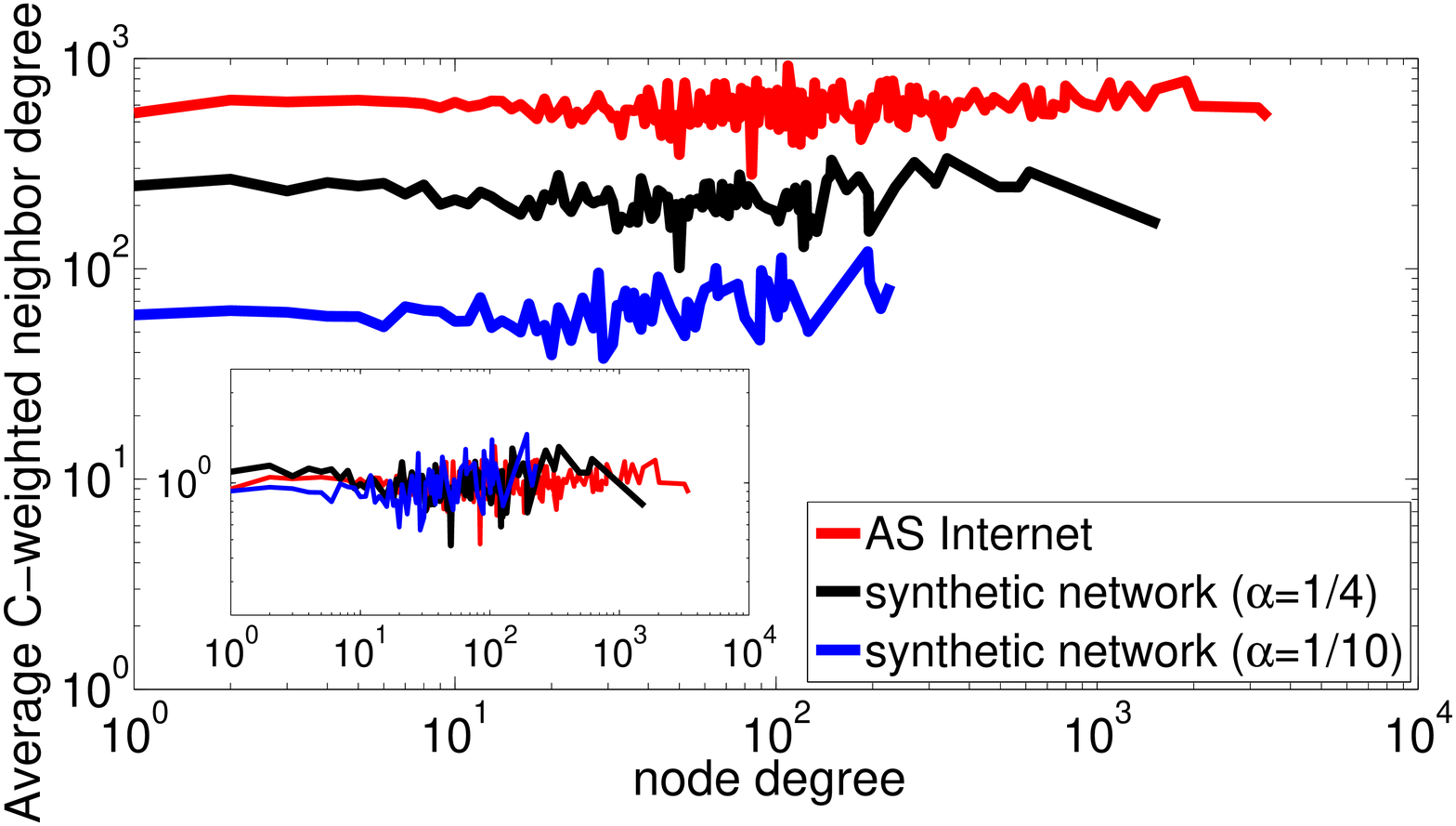}}\hfill
\subfigure[Propagation delay correlations.]{\includegraphics [width=1.8in, height=1.29in]{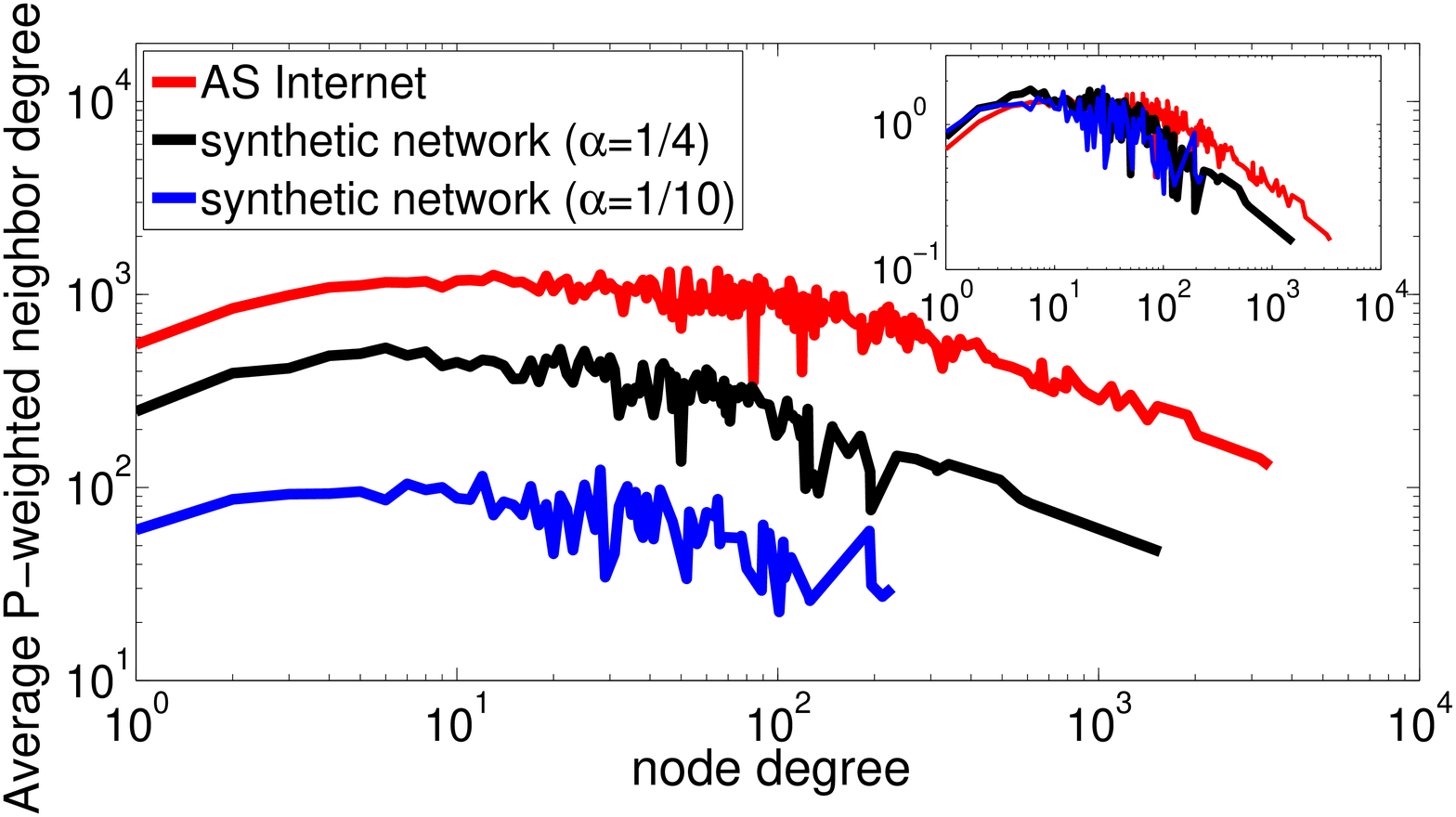}}\hfill
\subfigure[Normalized load vs. degree.]{\includegraphics [width=1.8in, height=1.3in]{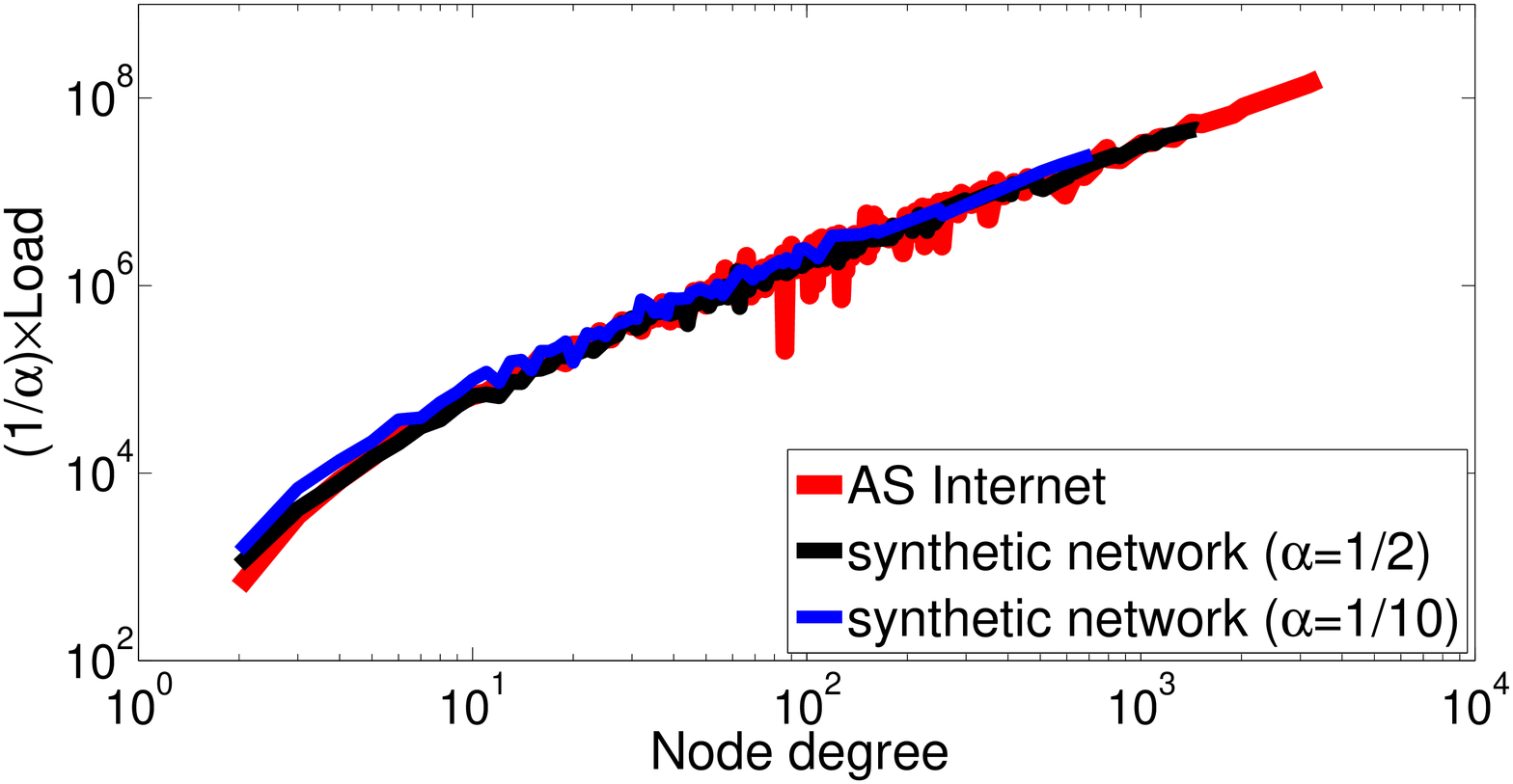}}\hfill
\subfigure[Distance distribution.]{\includegraphics [width=1.8in, height=1.3in]{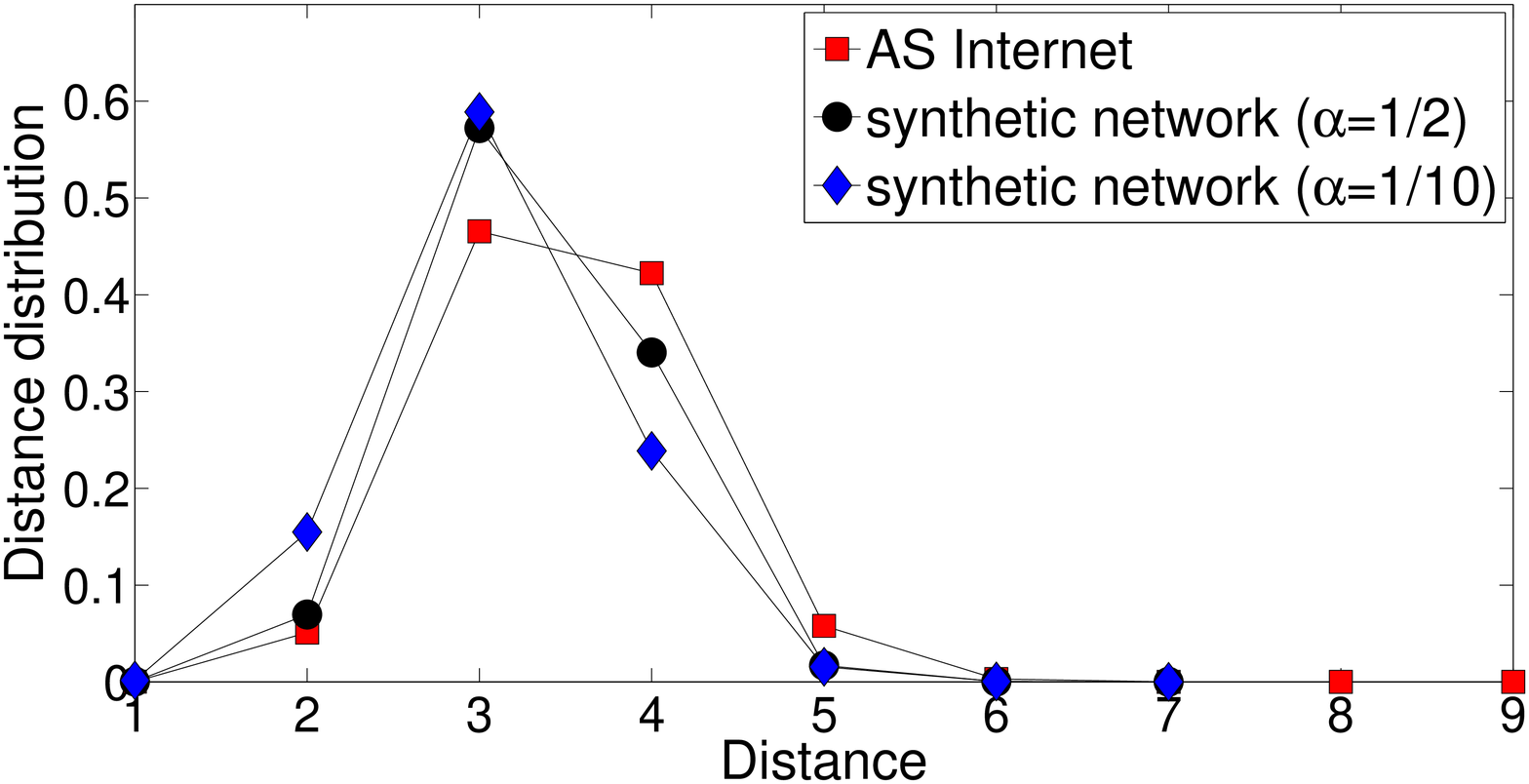}}}
\caption{Capacity and propagation delay correlations (plots~(a),(b)) in downscaled synthetic networks and the original AS Internet topology.  
The insets show the same curves normalized by their average value. Plots (c),(d) show the normalized load and the distance distribution.}
\label{fig:rescaled_vs_real_correlations}
\end{figure*}

The results for Scenario~1 are shown in Figures~\ref{fig:rescaled_vs_real_correlations}(a),(b). Specifically, let $C_{ij}$ be the capacity of the link connecting nodes $i, j$ and $P_{ij}$ be its propagation delay. For each node $i$ we compute its C-weighted neighbor degree, $k_C(k) =\frac{1}{s_i}\sum_{j=1}^{k} C_{ij}k_j$, where $k$ is the node's degree, $k_j$ is the degree of its $j^{th}$ neighbor, and $s_i=\sum_{j=1}^{k} C_{ij}$. Similarly, we compute its P-weighted neighbor degree, $k_P(k) =\frac{1}{s_i}\sum_{j=1}^{k} P_{ij}k_j$, where now $s_i=\sum_{j=1}^{k} P_{ij}$. Figure~\ref{fig:rescaled_vs_real_correlations}(a) shows the average C-weighted neighbor degree $\bar{k}_C(k)$, and Figure~\ref{fig:rescaled_vs_real_correlations}(b) shows the average P-weighted neighbor degree $\bar{k}_P(k)$. $\bar{k}_C(k)$ and $\bar{k}_P(k)$ are summary statistics, capturing correlations between degrees of connected nodes and link capacities or propagation delays, and are standard metrics used in  network theory~\cite{Barrat_pnas}. From Figures~\ref{fig:rescaled_vs_real_correlations}(a),(b), we observe that the shape of the curves remains the same, meaning that the capacity and propagation delay correlation characteristics are well preserved, even in replicas $10$ times smaller ($\alpha=1/10$) than the original network. Similar results hold for Scenario~2.

\begin{figure}
\centerline{
\subfigure[Degree distribution.]{\includegraphics [width=1.8in, height=1.3in]{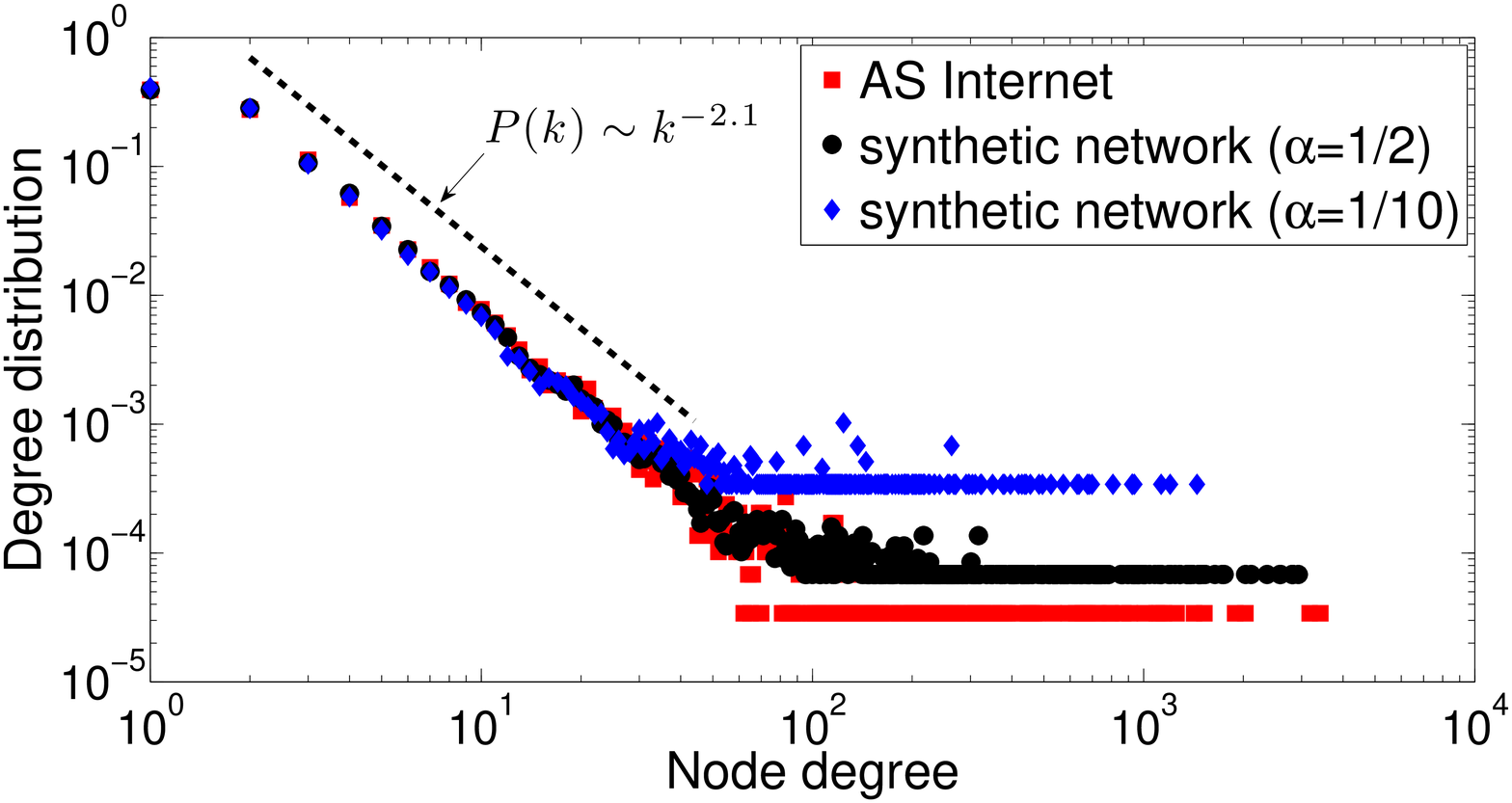}}\hfill
\subfigure[Clustering.]{\includegraphics [width=1.8in, height=1.3in]{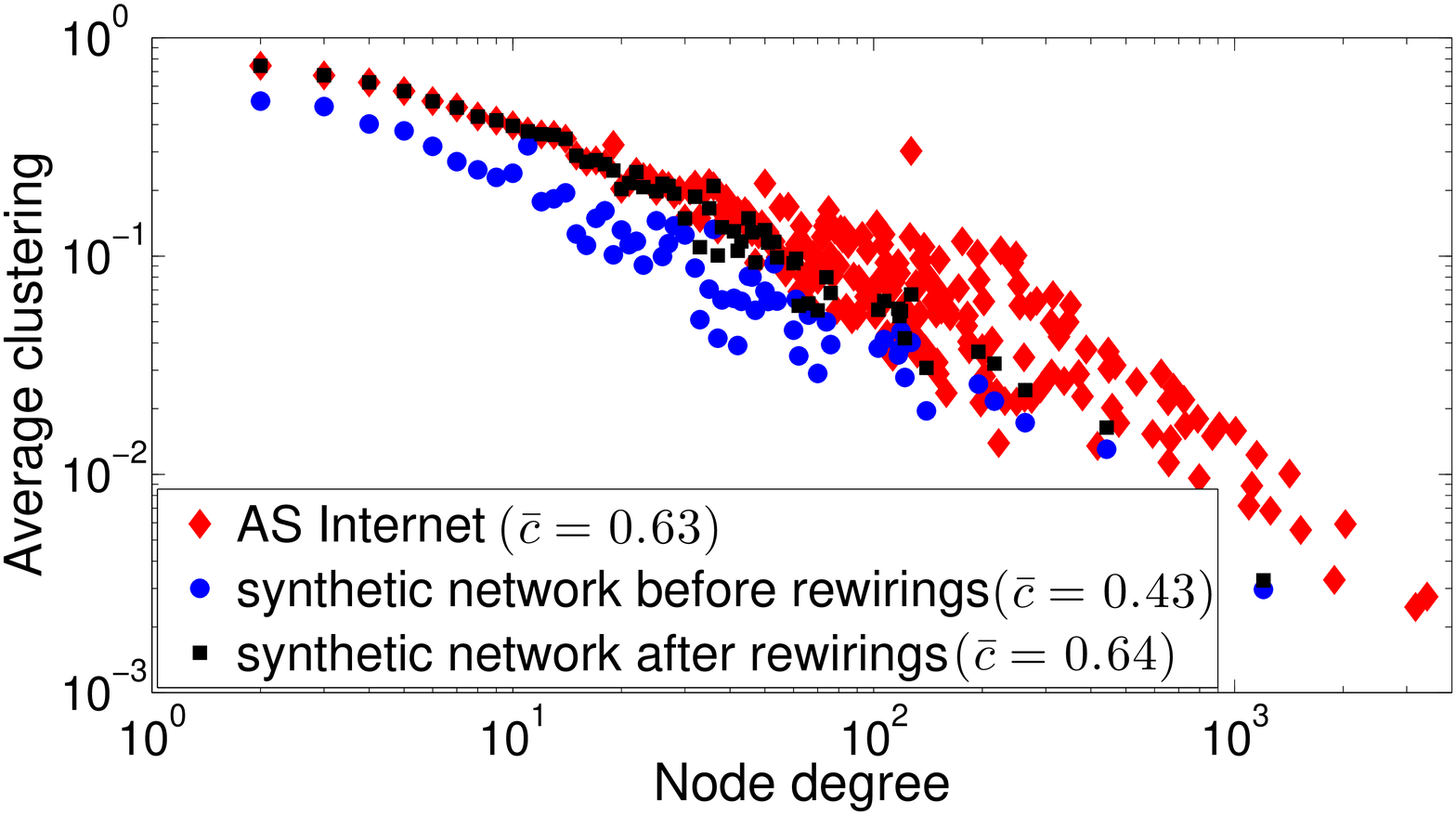}}}
\caption{Degree distribution and degree-dependent clustering.}
\label{fig:rescaled_vs_real_topology}
\end{figure}

Figure~\ref{fig:rescaled_vs_real_correlations}(c) shows the load $l(k)$ as a function of the node degree $k$, which is defined as
the average number of shortest paths in the topology passing through a $k$-degree node---this measure is also called node betweenness~\cite{Dorogovtsev10-book}. From the figure, we see that the \emph{normalized} load in the replicas matches well the load in the original network. This observation is in agreement with recent theoretical results~\cite{scaling_load}, showing that in scale-free networks of $N$ nodes and power law exponent $\gamma$,  $l(k) \sim N k^{\gamma-1}$, for $k \ll N^{\frac{1}{\gamma-1}}$. This observation also holds if we consider the load over network links instead of nodes. Furthermore, Figure~\ref{fig:rescaled_vs_real_correlations}(d) shows the distance distribution $d(h)$ in the three networks, i.e., the distribution of hop lengths $h$ of shortest paths between nodes. Its average value for the AS Internet and the two synthetic networks is $\bar{h}=3.496, 3.305, 3.11$, while its standard deviation is $\sigma_h=0.699, 0.624, 0.673$. We thus see that while the load on each node/link becomes smaller by the factor $\alpha$, path lengths change extremely slowly. Other topological properties of the original network are also approximately preserved in the smaller replicas.  For example $\bar{k}= 5.32, 5.05, 4.63$, while the assortativity and clustering coefficients~\cite{MaHu07} are $r=-0.18, -0.21, -0.26$, $c_{coeff}=0.016, 0.010, 0.018$, and the degree distribution $P(k)$ is shown in Figure~\ref{fig:rescaled_vs_real_topology}(a).
For each $\alpha$ these results represent averages over 10 generated graphs.

We note that since our method preserves the $2K$-distribution $p(k, k')$, it can reproduce the same topological properties as the studies in~\cite{MaHu07,tomacs_rescale}. However, similar to these studies, it cannot reproduce the amount of clustering in the original network, which requires preserving three-point degree correlations~\cite{MaKrFaVa06}. To address this issue,~\cite{MaKrFaVa06} and \cite{pol_clustering} suggest performing random link rewirings that preserve the $p(k,k')$ distribution and move clustering closer to that in the original network. Specifically, given two random links A--B and C--D whose end-points have degrees $(k_1, k_2)$ and ($k_3, k_4$), we rewire them to A--D and B--C given that $k_1=k_3$ or $k_2=k_4$, and that the rewiring moves clustering closer to the target clustering. Following this idea,~\cite{pol_clustering} showed that one can reproduce well the average clustering $\bar{c}(k)$ of $k$-degree nodes of the original network. This approach can be applied~\emph{as is} to replicas created by our procedure, since the aforementioned rewiring process does not alter the $p(k, k', C, P)$ distribution. Figure~\ref{fig:rescaled_vs_real_topology}(b) shows $\bar{c}(k)$, and reports the average clustering $\bar{c}$~\cite{MaHu07}, in the AS Internet and in a downscaled synthetic replica ($\alpha=1/10$) before and after applying the rewiring method of~\cite{pol_clustering}.

\section{Preserving Performance}
\label{sec:performance}

From a performance perspective, the downscaled replicas can be seen as networks that are ``approximately the same" with the original network, with the difference that the load (as defined in the previous section) on each link is reduced, i.e., multiplied, by the downscaling factor $\alpha$. Given this observation, performance can be preserved by utilizing the time-downscaling law from Section~\ref{sec:intro}.  Specifically, suppose that shortest path routing is used, and that flows between each possible source-destination pair $i,j$ arrive according to an independent Poisson process with rate $\lambda_{ij}$ drawn from some distribution. If the aggregate flow arrival process on a link connecting nodes of degrees $k, k'$ in the original network has rate $\lambda$, then in the downscaled replica it will have rate $\alpha \lambda$. (If flow $f_{ij}$ passes through link $l$, we call ``arrival time of $f_{ij}$ on $l$" the time that $f_{ij}$ arrived in the network.) This process is Poisson in both cases because it is the superposition of independent source-destination Poisson processes. Therefore, if we multiply all link capacities of the replica by $\alpha$ and divide all propagation delays and protocol timeouts by $\alpha$, then the time-downscaling law applies and performance is preserved.  Notice that here we do not sample flows as in \cite{shrink}---their arrival rate decreases by $\alpha$ because we reduce the network size by $\alpha$. Further, recall that packets within each flow can arrive according to \emph{any} process~\cite{shrink}. These arguments also hold, if instead of considering all possible source-destination pairs, we consider a random percentage $p$ of them in the original and downscaled networks.

\textbf{Validation.} To validate our arguments we consider the AS Internet topology from the previous section, and for each of the two scenarios described there (Scenarios 1, 2) we create small-scale replicas consisting of $N'=3000, 2000, 1000, 500$ nodes. We use these topologies in the ns-3 simulator~\cite{ns3}, after scaling their capacities and propagation delays by the factor $\alpha=1, 2/3, 1/3, 1/6$, respectively, as previously described. In each topology, we randomly select a percentage $p=10\%$ of source-destination pairs. Between each selected pair, flows arrive according to a Poisson process with rate $\lambda=0.1$ flows/sec for Scenario 1 and $\lambda=0.05$ flows/sec for Scenario 2. Each flow consists of a Pareto-distributed number of packets, with an average size of $4$ packets, maximum size of $10^{4}$ packets, and a shape parameter equal to $1.2$. In Scenario 1 the flows are TCP, while in Scenario 2 the flows are UDP. In an $\alpha$-scaled replica the timeouts of the TCP flows are divided by the factor $\alpha$ (which is accomplished by dividing by $\alpha$ the TCP's initial round trip time estimate and its minimum retransmission timeout), while UDP flows transmit at a constant rate of $6\alpha$ packets/sec. The packet size is $1000$ bytes, and the buffers at the nodes use DropTail and can hold $300$ packets. The simulation time for an $\alpha$-scaled replica is $100/\alpha$ seconds.  

\begin{figure}
\centerline{
\subfigure[Flow completion time.]{\includegraphics [width=1.8in, height=1.2in]{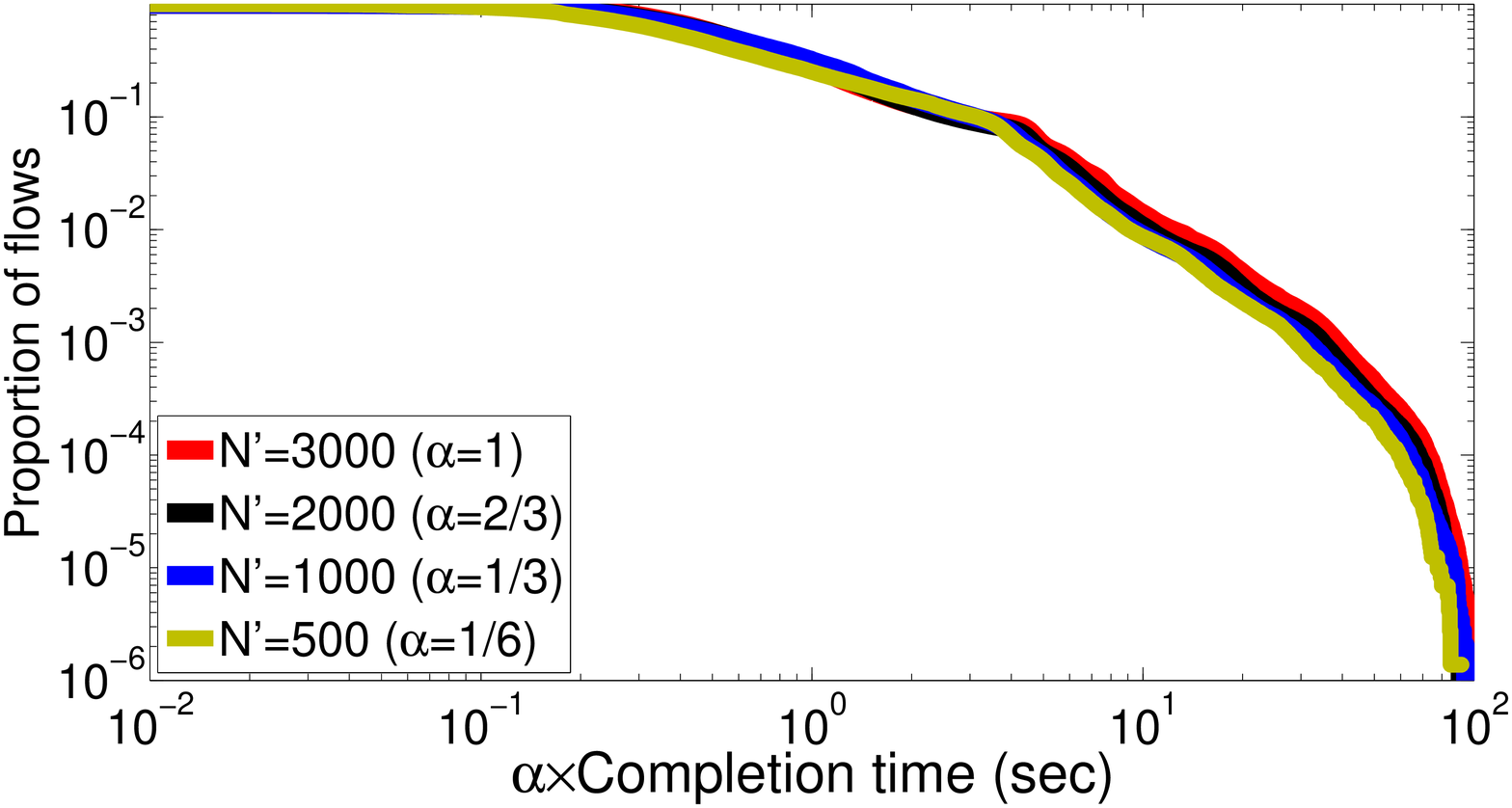}}\hfill
\subfigure[End-to-end packet delay.]{\includegraphics [width=1.8in, height=1.2in]{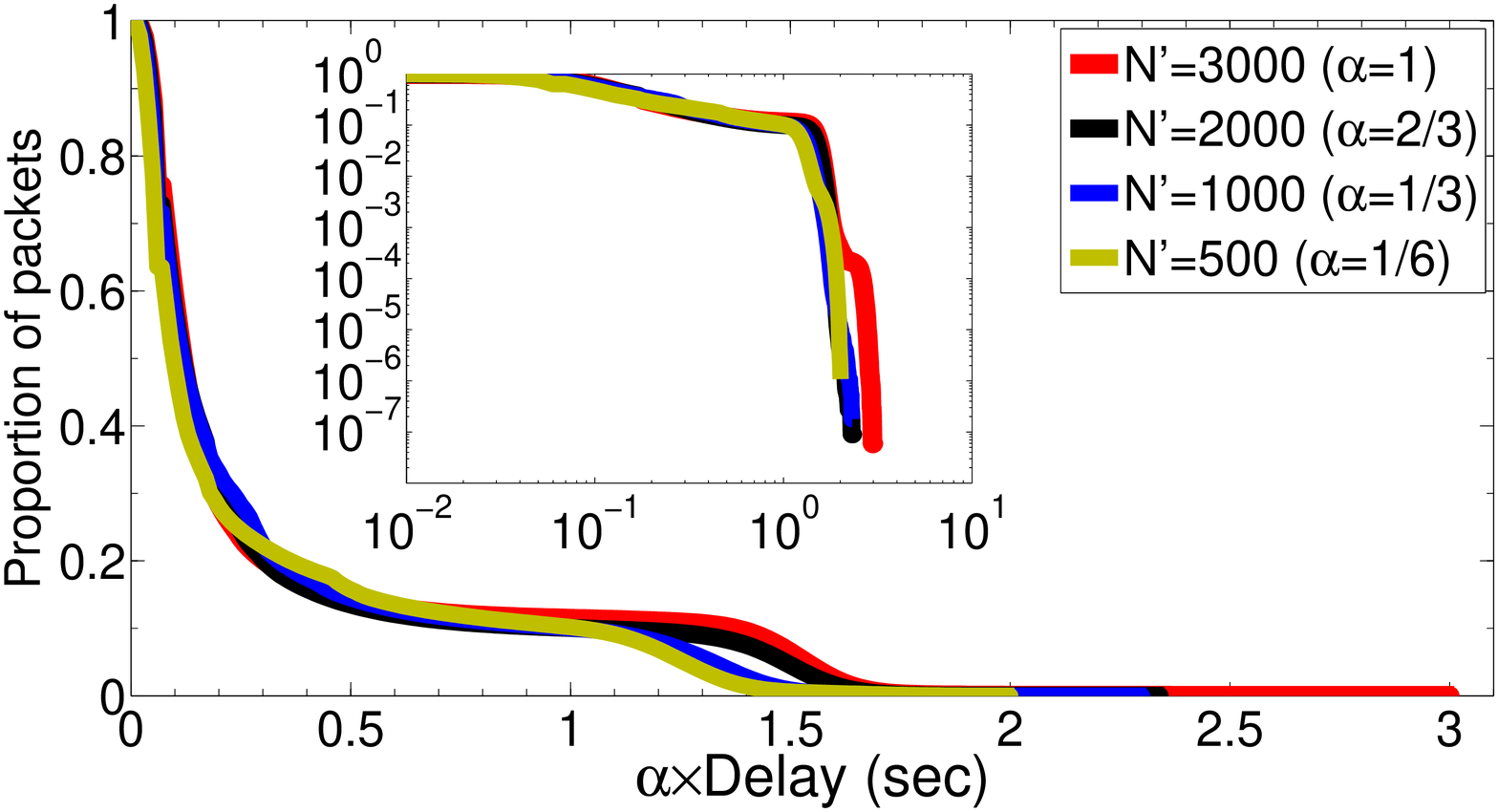}}}
\centerline{
\subfigure[Flow completion time.]{\includegraphics [width=1.8in, height=1.2in]{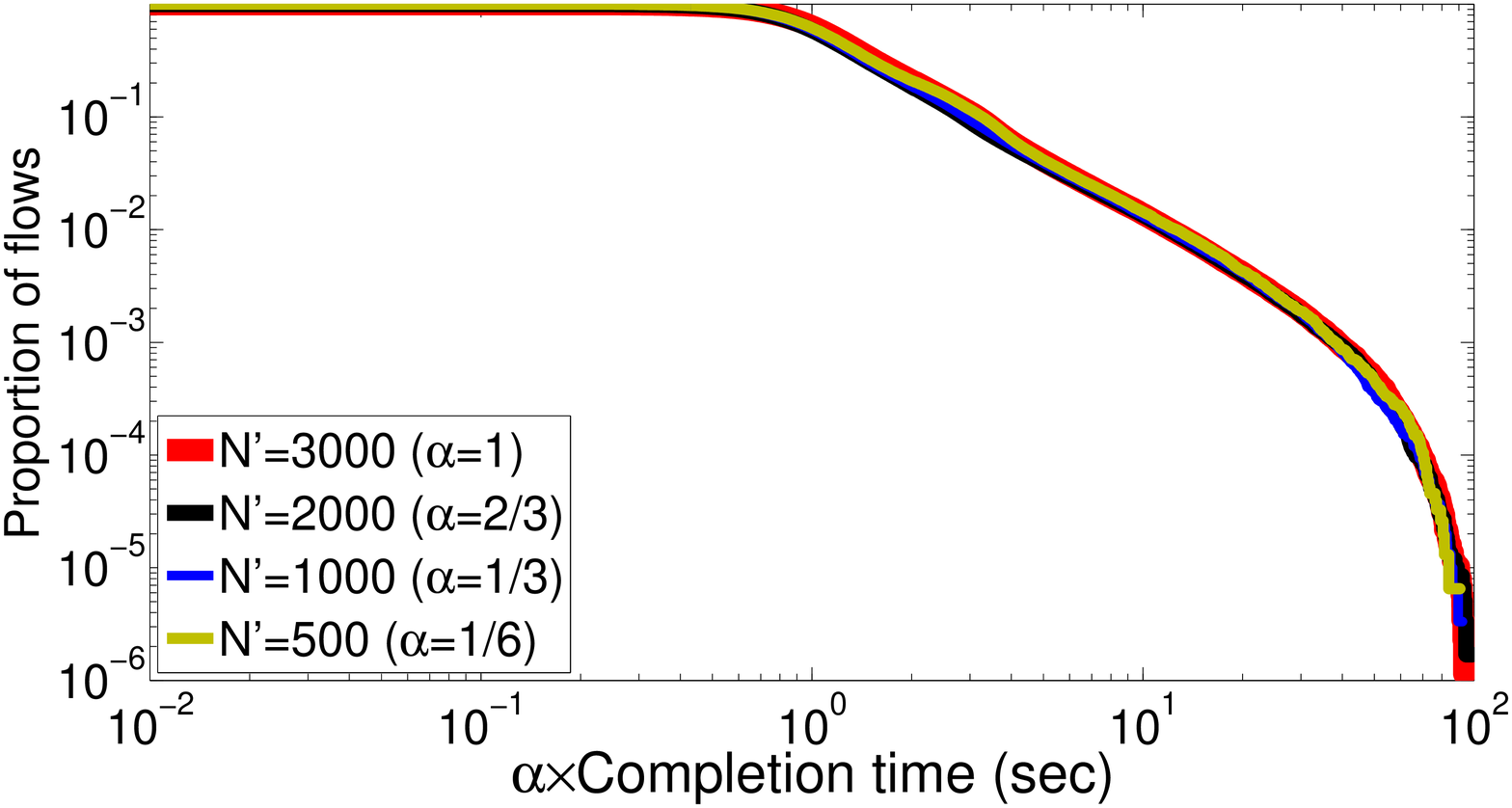}}\hfill
\subfigure[End-to-end packet delay.]{\includegraphics [width=1.8in, height=1.2in]{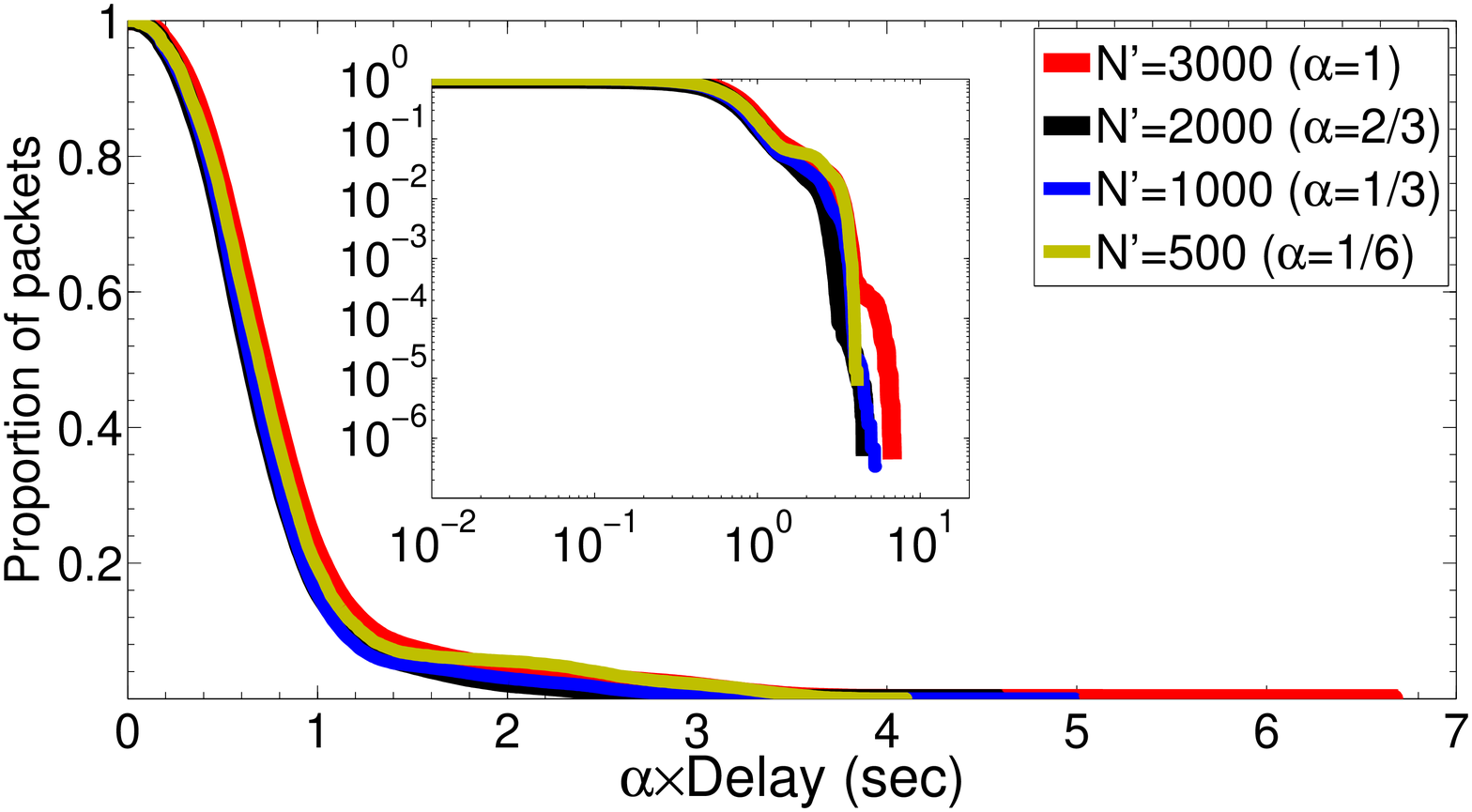}}}
\caption{CCDF of the normalized flow completion time and end-to-end packet delay. Plots (a),(b) correspond to Scenario 1 and plots (c),(d) to Scenario 2. The insets in plots (b),(d) show the same distributions in log-log scale.}
\label{fig:performance}
\end{figure}

Figures~\ref{fig:performance}(a),(c) show, in log-log scale, the distribution of the normalized flow completion time in the four network replicas of each scenario. The flow completion time is defined as the time elapsed from the moment that the first packet of a flow arrives in the network till the moment that the last packet of the flow departs the network. From the figures, we observe a remarkable match in the distributions, in agreement with our theoretical arguments. Similar observations hold in Figures~\ref{fig:performance}(b),(d), which show the distribution of the normalized end-to-end packet delay, that is, the distribution of the total delay that a packet experiences from the moment the packet enters the network till the moment the packet departs the network. From Figures~\ref{fig:performance}(b),(d), we observe a very good match in the distributions, with only small differences at their tails involving a very small percentage of packets. These differences are  expected for two reasons: first, a small percentage of packets can traverse slightly longer paths as the network size increases, see Figure~\ref{fig:rescaled_vs_real_correlations}(d) and the related discussion; and second, in larger networks some packets can traverse higher degree nodes, which do not exist in the smaller networks (since $k_{max} \sim N^{\frac{1}{\gamma-1}}$, see Fig.~\ref{fig:rescaled_vs_real_topology}(a)), and which may have low capacity links attached to them, depending on the scenario. However, as it is evident by Figures~\ref{fig:performance}(a-d), these differences do not have a significant contribution to the overall performance of the replicas.

To highlight the practical benefits of our approach, we report the time needed for the simulations to complete in each of the considered scenarios.  Scenario~1 with $N'=3000, 2000, 1000, 500$ nodes required respectively $25$ days, $8$ days, $2$ days and $9$ hours, while Scenario~2 required $5$ days, $2$ days, $11$ hours and $3$ hours. We see that the decrease in the simulation time with the size of the network is astonishing.  All simulations were run using a CPU with speed $\sim 3$ GHz.
The TCP simulations (Scenario 1) were the most computationally demanding, requiring $68$ GB of memory (RAM) when $N'=3000$, while only $4$ GB RAM were needed for $N'=500$.  The UDP simulations (Scenario 2) required $14$ GB RAM for $N'=3000$, and only $1$ GB RAM for $N'=500$.

\section{Conclusion}
\label{sec:conclusion}

Our results show that it is possible to efficiently and accurately predict the performance of large complex networks, using suitably scaled-down replicas consisting of a significantly smaller number of nodes. To our best knowledge, this is the first time that this has been demonstrated, and supported by theory. There are several interesting directions for future research. One is to apply our procedure to other power-law topologies.  Another is to investigate whether similar results hold when flow arrivals are not dictated by Poisson processes. Finally, our work proposes a general procedure to construct correlated weighted networks based on input from real data, and could also find applications in other contexts~\cite{Barrat_pnas}.

\bibliographystyle{IEEEtran}
\balance
\bibliography{bib}

\end{document}